\documentclass[runningheads,fleqn]{svmult}
\usepackage{makeidx}   
\usepackage{graphicx}  
\usepackage{subeqnar}  
\usepackage{multicol}  
\usepackage{taphys}    
\makeindex             
\usepackage[]{amsbsy}
\begin{document}
\title*{Metal-Insulator Transitions at Surfaces}
\toctitle{Metal-Insulator Transitions at Surfaces}
%
%
\titlerunning{Metal-Insulator Transitions at Surfaces}
%
\author{Michael Potthoff}
\authorrunning{Michael Potthoff}
%
%
\institute{Lehrstuhl Festk\"orpertheorie, Institut f\"ur Physik,
Humboldt-Universit\"at zu Berlin, Germany}

\maketitle              

\begin{abstract}
Various types of metal-insulator transitions are discussed to find 
conditions for which an ideal surface of a bulk insulator is metallic.
It is argued that for the correlation-driven Mott metal-insulator transition
the surface phase diagram should be expected to have the same topology 
as the phase diagram for magnetic order at surfaces:
The corresponding linearized mean-field descriptions, a simplified dynamical 
mean-field theory of the Hubbard model and the Weiss mean-field theory for 
the Ising model, are found to be formally equivalent.
A new kind of surface state appears in the low-energy part of
the one-particle excitation spectrum as a precursor effect of the
Mott transition.
\end{abstract}

The Mott metal-insulator transition at a crystal surface is a 
subject that touches different areas in solid-state theory which are 
usually treated as being disjoined:
many-body theory of the correlation-driven metal-insulator transition,
the general theory of surface phase transitions, and the theory of 
electronic surface states.
It is the intention of the present paper to show that a corresponding 
combination of different concepts can be fruitful and allows for some 
new theoretical predictions.

\section{Surface Phase Transitions}

The large variety of novel and interesting phenomena in surface physics
is closely related with the occurrence of surface phase transitions.
As has been pointed out by Mills \cite{Mil71},
the surface of a system may undergo a phase transition 
at a critical temperature $T_{\rm c,s}$ being different from the bulk 
critical temperature $T_{\rm c}$, i.e.\
the surface may undergo its own phase transition.
Critical exponents, for example, can be defined and determined which 
are specific for the transition at the surface and which cannot be 
fully reduced to the bulk critical exponents \cite{Bin83,Die86}.
Different kinds of surface phase transitions are conceivable
and have been found, e.g.\ structural transitions, such as deviating 
geometrical order of the atoms near the surface of a single crystal 
(surface reconstruction), the loss of long-range crystalline order at the 
surface prior to a bulk melting transition (surface melting) 
or enrichment of one component at the surface of a solid binary alloy 
(surface segregation) \cite{Bin83}.
Typical examples for surface phase transitions are also found among 
magnetically ordered systems:
For example, the (0001) surface of ferromagnetic Gd is believed to have a 
Curie temperature which is higher than the bulk $T_{\rm C}$ \cite{DDN98}. 

Different types of surface phase transitions can be described in a 
qualitative but consistent way by means of classical Landau theory 
\cite{Bin83,Die86} or by lattice mean-field approaches which may be 
considered as coarse-grained realizations of the Landau theory.
Especially, mean-field approaches to localized-spin models, such as the
Ising or Heisenberg model, are frequently considered in this context 
\cite{isihei}.
For surface geometries there are a number of non-trivial results predicted 
by Landau or mean-field theory, such as temperature-dependent order-parameter 
profiles, which may give a surprisingly good description of experimental data 
(see Ref.\ \cite{PP01}, for example).

Typically, the surface undergoes the phase transition at the same 
temperature as in the bulk, $T_{\rm c,s} = T_{\rm c}$, if the local 
(structural, electronic, magnetic) environment remains unchanged, while 
$T_{\rm c,s} > T_{\rm c}$ if there is a perturbation $\Delta$ at the 
surface exceeding a certain critical value $\Delta_{\rm c}$.
For $\Delta > \Delta_{\rm c}$ and temperatures $T_{\rm c} < T < T_{\rm c, s}$
there is an ordered $D-1$-dimensional surface coexisting with a disordered
$D$-dimensional bulk.
More complicated phase diagrams are obtained in the case of multi-critical
behavior, e.g.\ when the long-range order at the surface has a character
different from the long-range order in the bulk (surface reconstruction,
anti-ferromagnetic surface of a ferromagnetic bulk, etc).
The Landau $T$--$\Delta$ phase diagram should be qualitatively correct
whenever the $D-1$-dimensional system can support independent order 
\cite{Bin83,MW66}.

\section{Metal-Insulator Transitions}

The concept of a surface phase transition and the corresponding Landau 
theory seems to extend straightforwardly to a certain kind of metal-insulator
transitions, namely those which accompany an order-disorder thermodynamic 
phase transition (see Ref.\ \cite{Geb97} for an overview).
The thermodynamic phase transition will be considered at the $T=0$ quantum-critical
point as due to ubiquitous thermal activation processes, a strict definition of a 
metal-insulator transition is possible at zero temperature only.
It is well known that the formation of an ordered state may result 
in a gap for charge excitations as, for example, in the Peierls transition or 
in the transition to an anti-ferromagnetic state:
Consider the typical example of a bipartite lattice and two-sublattice 
long-range order causing a doubling of the unit cell.
For a non-degenerate band the splitting at the boundary of the new Brillouin 
zone will lead to a gap and, in the case of half-filling, to an insulating 
ground state.

Now, for sublattice order at the surface of a disordered bulk one would 
have the (naive) expectation that an insulating surface could coexist with 
a metallic bulk.
This, however, is clearly impossible as a finite bulk density of states at 
the Fermi energy will always induce a non-zero, though possibly low density 
of states in the surface region.
Likewise, the reverse scenario is impossible either:
Namely, to realize a metallic surface phase of a bulk insulator caused by a 
thermodynamic phase transition, a disordered surface would have to coexist 
with an ordered bulk which, in general, is forbidden by strict arguments 
\cite{Bin83}
(though a magnetic ``dead-layer'' scenario is found under somewhat 
exotic circumstances in a $D=2$ $q$-state surface Potts model 
\cite{Lip}).

Besides the thermodynamic transitions, there is a second important class of 
metal-insulator transitions, namely quantum-phase transitions \cite{Geb97}.
Essentially these take place at $T=0$ only and are not associated with any
symmetry breaking.
Important examples are the transition from a metal to a normal band insulator
and the Mott-Hubbard transition from a metallic Fermi liquid to a Mott 
insulator.
While the former can be understood within an independent-electron model, 
correlation effects are constitutive for the latter.
Referring to a quantum-phase metal-insulator transition, it is very well 
feasible that the surface of a bulk insulator is metallic.

\section{Surface States}

\begin{figure}[t]
\centerline{\includegraphics[width=.4\textwidth]{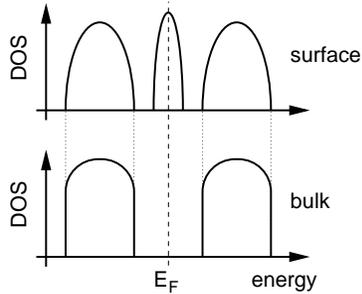}}
\caption[]{
Bulk and the surface density of states (schematic) in case of a 
metallic surface and an insulating bulk. $E_{\rm F}$: Fermi energy.}
\label{fig:schema}
\end{figure}

Fig.~\ref{fig:schema} shows a possible electron density of states for this 
situation. 
To have a metallic surface of an insulator, the density of states at 
the Fermi energy $E_{\rm F}$ must be finite at the surface while, with
respect to the bulk states, $E_{\rm F}$ should lie within a band gap.
Note that this necessarily implies the existence of a partially filled 
surface state:
The appearance of a surface state at the Fermi energy is crucial 
to get a metallic surface of an insulator.

Two possible origins of electronic surface states are well known \cite{DS92}:
(i) Image-potential states may arise as Rydberg-like states in the 
long-ranged $-1/4z$ image potential which is due to the polarization
charge that is induced by an electron approaching a {\em conductive} 
surface. 
The electron can be trapped between the image-potential surface barrier 
and the bulk barrier due to a bulk band gap. 
(ii) Crystal-induced surface states originate from the mere crystal 
termination.
For an ideal unreconstructed crystal surface a further distinction 
between Tamm and Shockley states is meaningful:
Shockley states appear within a hybridizational band gap which may open 
when the boundaries of two bulk bands have crossed as a function of 
decreasing lattice constant.
Tamm-like surface states are due to the surface change of the one-electron 
potential and always lie near the bulk band from which they originate. 


In fact, a crystal-induced surface state may give rise to a metallic surface 
phase as is demonstrated by the following two examples:
Due the reconstruction of the Si(111)-$(7\times 7)$ surface there is 
a partially occupied surface state which is consistent with the observed 
metallic behavior for this surface (cf.\ the discussion in Ref.\ \cite{DS92}).
A surface insulator-to-metal transition has been predicted for the 
ferromagnetic insulator EuO \cite{SN01}:
For temperatures below $T_{\rm C}= 69$~K and decreasing, the majority $5d$ 
conduction band shifts towards the occupied $4f$-$\uparrow$ bands thereby 
reducing the insulating gap.
This so-called red shift is transmitted to an unoccupied Tamm state which 
is predicted to split off at the (100) surface from the lower edge of the 
conduction band.
The energy difference between the surface state and the majority bulk band 
edge together with the $T$-dependent red shift is just sufficient to bridge 
the gap.
Therefore, the surface state should become populated for low $T$ which would
imply a transition to a (half-)metallic state at the surface.

In any case the concept of a crystal-induced surface state is based on a 
model of effectively independent electrons. 
It is the detailed form of the one-electron potential that determines the 
energy position (lateral dispersion) of a surface state in the band structure.
Hence, it is obvious that a surface state is not normally pinned to 
$E_{\rm F}$ (chemical potential); 
one may conclude that for a surface of a band insulator there is no 
{\em a priori} reason for a surface state to be partially filled.

\section{Mott Transition}

The question is whether or not a situation is conceivable where the appearance 
of the surface state is necessarily connected with the metal-insulator
transition, i.e.\ where the surface state necessarily appears at $E_{\rm F}$.
It is now clear that to this end one has to look for an electron-correlation 
effect.

Consider electrons in a narrow non-degenerate band interacting with each 
other via an on-site Coulomb repulsion $U$ as described by the 
Hubbard model \cite{Hub}.
Any symmetry-broken phases will be excluded from the discussion.
At half-filling $n=1$ and for strong $U$, the system is then a paramagnetic 
insulator.
The ${\bf k}$-integrated one-electron excitation spectrum (DOS) of this 
so-called Mott-Hubbard insulator \cite{Geb97,Mot90} has the same form as 
shown in 
the lower part of Fig.~\ref{fig:schema} where now the two peaks have to be 
interpreted as the lower and the upper Hubbard band separated by an energy 
of the order of $U$.

On the other hand, for $U=0$ and for the weakly interacting case, the system
is a normal Fermi liquid. 
At an intermediate interaction strength $U_{\rm c}$ of the order of the 
width $W$ of the non-interacting DOS one thus expects a metal-insulator 
transition.
This Mott transition is a prime example for a quantum phase transition at
$T=0$ which results from the competition between the electrons' kinetic
energy $\sim W$ which tends to delocalize the electrons and their potential 
energy $\sim U$ which tends to localize them.

The Mott transition is a true many-body effect that cannot be explained 
by a simple perturbational approach.
Even within the framework of simplified model Hamiltonians, as the
Hubbard model, an ultimate theory of the Mott transition is still missing
\cite{Geb97}.
A decisive step forward, however, has been made in the last decade with 
the development of dynamical mean-field theory (DMFT) and its application 
to the Mott transition (see Sec.\ \ref{sec:mfa}, for a review see Ref.\ 
\cite{GKKR96}, recent results can be found in Ref.\ \cite{BCVJO}).

Within the DMFT one finds that the transition at $U=U_{\rm c}$ is 
characterized by a diverging effective mass $m^\ast \to \infty$ or 
a vanishing quasi-particle weight $z \propto (m^\ast)^{-1} \to 0$, 
respectively.
For $U < U_{\rm c}$ but close to the critical point, the DOS has a 
three-peak structure, consisting of the two well-developed Hubbard bands 
as well as a narrow quasi-particle resonance at the Fermi energy with
weight $z$ -- the DOS has the same form as the ``surface DOS'' shown by 
the upper part of Fig.~\ref{fig:schema}.

It is conceivable that Fig.~\ref{fig:schema} describes a situation where 
the bulk of the system is a Mott-Hubbard insulator while the surface is in 
a metallic Fermi-liquid state.
The quasi-particle resonance would then be a surface state (one-electron
surface excitation) with a layer-dependent weight $z_\alpha$ decreasing 
exponentially with increasing distance from the surface.
This surface state necessarily appears at the Fermi energy as it 
corresponds to low-energy excitations well known from quantum impurity systems
(Kondo effect).
The question is for which circumstances this coexistence of the Mott-Hubbard 
insulator and the metallic Fermi liquid can be realized.

\section{Mean-Field Approach}
\label{sec:mfa}

As it is by no means obvious how to construct a (continuum) Landau theory for 
this problem, the method of choice is to formulate and evaluate a mean-field 
theory for an appropriate discrete lattice model.
While for a magnetic phase transition one can resort to effective spin models 
such as the Ising model without any detailed knowledge of the electronic 
structure, the Hubbard model as the minimum model to describe the Mott 
transition includes spin as well as charge degrees of freedom and is thus 
much more complicated.
Likewise it is much more complicated to find a proper mean-field theory.
For example, Hartree-Fock theory, weak- and strong-coupling perturbational 
approaches or decoupling approximations for the Hubbard model are clearly 
inferior compared with the Weiss mean-field theory for the Ising model.
The latter is non-perturbative, thermodynamically consistent and free of
unphysical singularities in the entire parameter space.
Since the Weiss theory becomes exact in the non-trivial limit of infinite 
spatial dimensions $D$ \cite{Eng63}, this may serve as a simple and precise 
characterization of what is a proper mean-field theory in general.
One may therefore hope that the same limit will lead to a powerful mean-field
approach in the case of the Hubbard model, too.

That the $D=\infty$ limit for a lattice fermion model is well-defined and 
non-trivial in fact, has been proven in the pioneering work of Metzner and 
Vollhardt \cite{MV89}.
Crucial is a proper scaling of the hopping $t \propto 1 / \sqrt{D}$ to 
keep the dynamic balance between kinetic and potential energy.
To convert the abstract definition of a proper mean-field theory into a
useful concept for practical calculations, it has been important to realize
that the $D=\infty$ Hubbard model can be mapped onto the single-impurity 
Anderson model (SIAM) as now one can profit from various methods available 
for impurity problems.
This observation has been made by Georges, Kotliar and Jarrell 
\cite{GK92,Jar92}.
The mapping is a self-consistent one which means that the parameters of
SIAM depend on the one-particle Green function of the Hubbard model.

To study surface effects one has to consider a variant of the original 
Hubbard model. 
Using standard notations the Hamiltonian reads:
\begin{equation}
  H = \sum_{i_\| j_\| \alpha \beta \sigma} 
  t_{i_\|\alpha,j_\|\beta} \:
  c^\dagger_{i_\|\alpha\sigma} c_{j_\|\beta\sigma}
  + \frac{U}{2} \sum_{i_\|\alpha\sigma} \: 
  n_{i_\|\alpha\sigma} n_{i_\|\alpha -\sigma} \: .
\label{eq:hubbard}
\end{equation}
Here $i_\|$ labels the sites within a layer $\alpha$ parallel to the 
surface. 
$\alpha=1$ corresponds to the top surface layer.
The model (\ref{eq:hubbard}) differs from the original Hubbard model by 
the mere existence of the surface: The sole effect of the surface is to 
terminate the bulk. 
For any numerical calculation one has to assume a finite number of layers,
$\alpha = 1,...,d$, i.e.\ a film geometry, and to check the convergence
of the results for large $d \to \infty$.

\begin{figure}[t]
\mbox{} \hfill
\raisebox{5mm}{\includegraphics[width=0.45\textwidth]{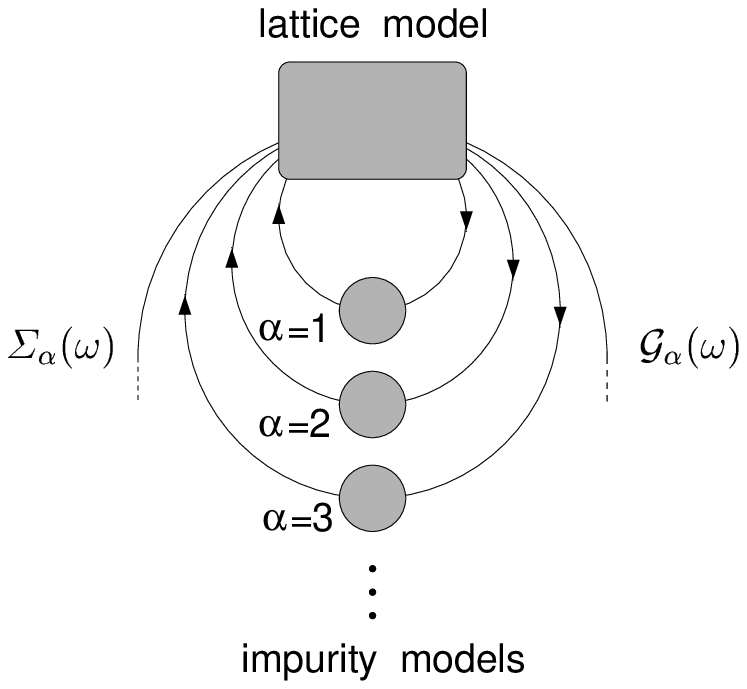}}
\hfill
\includegraphics[width=.37\textwidth]{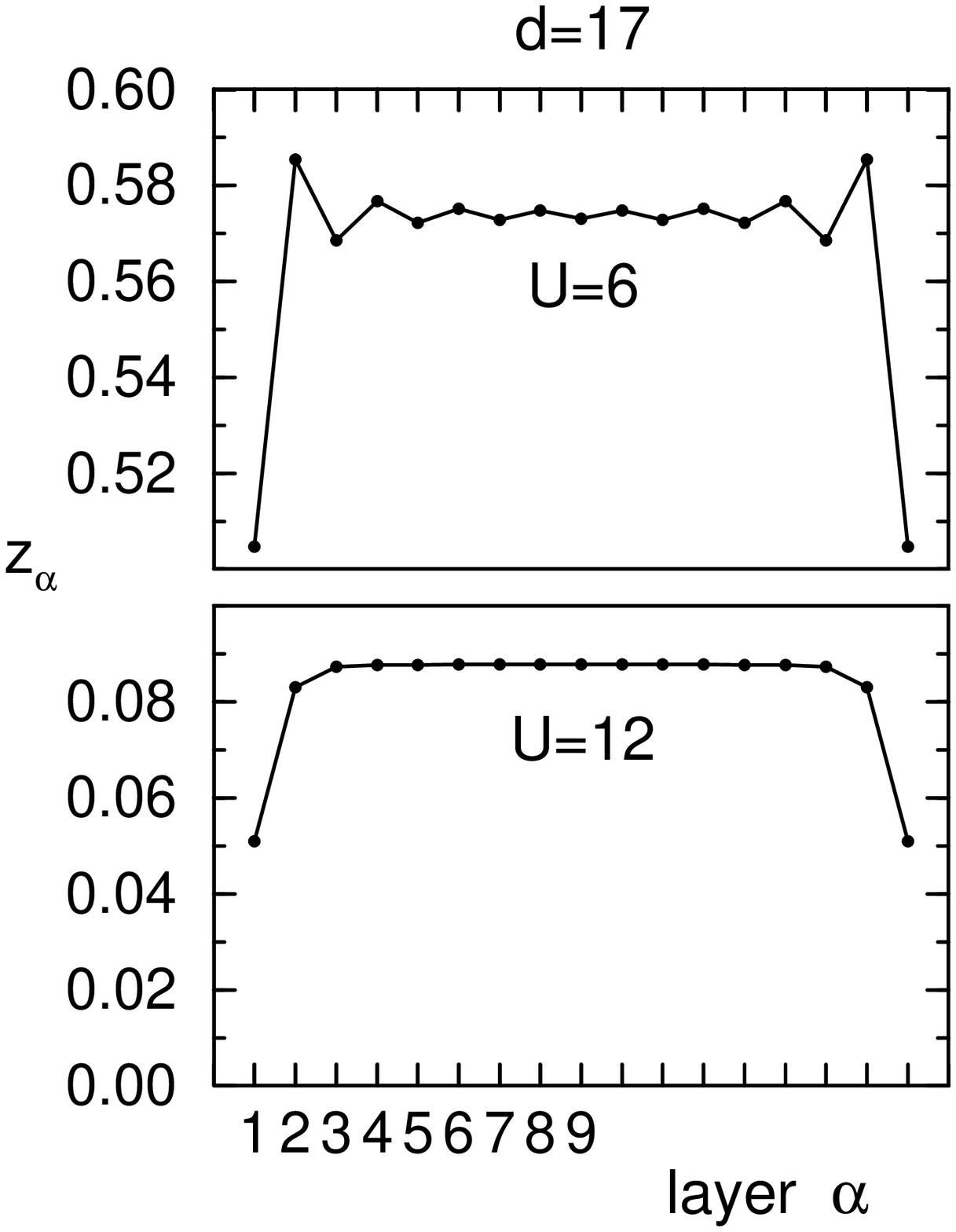}
\hfill
\mbox{}
\caption[]{{\em Left:}
           DMFT self-consistency cycle for the Hubbard model on a
           semi-infinite lattice with layer index $\alpha=1,...,d$ 
	   ($d\to\infty$). See text for discussion.
	   {\em Right:}
	   Layer-dependent quasi-particle weight $z_\alpha$ for a 
	   simple-cubic film of $d=17$ layers with (100) surfaces. 
	   DMFT results for uniform nearest-neighbor hopping $t=1$, 
	   half-filling $n=1$, $T=0$ and $U$ as indicated.}
\label{fig:dmft}
\end{figure}

The generalization of DMFT for surface geometries has been developed
by Potthoff and Nolting \cite{PN}.
Again, the limit $D=\infty$ may serve as a guide to construct a powerful
mean-field theory.
Assuming uniform hopping parameters, $t_{i_\|\alpha,j_\|\beta} = t$ between 
nearest neighbors $(i_\|,\alpha)$ and $(j_\|,\beta)$, and using the same 
scaling $t \propto 1 / \sqrt{D}$, the model itself as well as surface 
effects remain non-trivial for $D \to \infty$.
As the different layers parallel to the surface must be treated as being 
inequivalent, the mapping procedure, however, becomes more complicated 
(see Fig.\ \ref{fig:dmft}, left).
The original many-body problem for a semi-infinite lattice with layers
$\alpha=1,2,...,d$ (with $d \to \infty$) is self-consistently mapped onto a 
{\em set} of impurity problems labeled by the same index $\alpha=1,2,...,d$.
Each SIAM can be treated independently to calculate the impurity self-energy 
$\Sigma_\alpha(\omega)$. 
There is, however, an indirect coupling which is mediated by the 
self-consistency cycle: 
Via the Dyson equation of the lattice model, the on-site Green function 
$G_\alpha(\omega)$ for a given layer $\alpha$ depends on {\em all} 
layer-dependent self-energies. 
The free Green function ${\cal G}_\alpha(\omega)$ of the $\alpha$th SIAM 
which determines its one-particle parameters is then obtained from the DMFT
self-consistency condition:
${\cal G}_\alpha(\omega)^{-1} = G_\alpha(\omega)^{-1} + \Sigma_\alpha(\omega)$.

Fig.\ \ref{fig:dmft} (right) shows the quasi-particle weight $z_\alpha = 
(1 - \Sigma'_\alpha(0))^{-1}$ as obtained from the DMFT using a standard 
(``exact diagonalization'') method \cite{CK94} to treat the different 
impurity problems.
The critical interaction $U_{\rm c}(d)$ of the $d=17$ layer sc(100) film 
lies close to the bulk critical interaction $U_{\rm c, bulk} \approx 16$ ($W=12$ 
is the width of the free bulk DOS).
In the metallic phase for $U < U_{\rm c}(d)$ there is a quasi-particle 
resonance in the interacting local density of states for each layer with
a finite weight $z_\alpha$.
As can be seen in the figure, the weight has a pronounced layer dependence.
While for small $U$ the profile has an oscillating character, it becomes 
monotonous for interaction strengths close to the transition.
This is a typical result which is observed for films with different surface
geometries and indicates a universal behavior of the critical profile for
$U \to U_{\rm c}(d)$.
For both, $U=6$ and $U=12$, the surface quasi-particle weight $z_1$ is
considerably lower than the bulk quasi-particle weight $z$ at the film
center.
This result is plausible:
Due to the reduced coordination number at the surface $q_1 < q$, the 
variance $\Delta_1 = q_1 t^2$ of the surface-layer DOS is reduced which
implies the ``effective'' interaction $U / \sqrt{\Delta_1}$ to be stronger
at the surface compared with the bulk.
In this respect the surface is ``closer'' to the insulating phase.
Yet, for $U \to U_{\rm c}(d)$ all $z_\alpha$ vanish simultaneously
and there is no surface transition.

\section{Critical Regime}

For systematic investigations of films with different (large) thicknesses,
with different surface geometries and for different model parameters, a 
numerically exact evaluation of the DMFT requires an effort which is out 
of scale.
Fortunately, a simplified treatment of the mean-field equations is possible
at $T=0$ for parameters close to the critical point as has been pointed out 
by Bulla and Potthoff \cite{BP00}.
This ``linearized DMFT'' (L-DMFT) is based on two plausible assumptions for
the critical regime $U \to U_{\rm c}$:
(i) The effect of the two Hubbard bands on the quasi-particle resonance 
can be disregarded and the resonance basically reproduces itself in the 
DMFT self-consistency cycle.
(ii) The resonance has no internal structure and can be described by a
one-pole approximation.
Although these assumptions are approximate, the L-DMFT has successfully 
passed a number of tests which have been performed by comparing with the 
full theory and which show that the L-DMFT is well qualified to give 
quantitative estimates for critical interactions and critical profiles
as well as the correct topology of phase diagrams
\cite{BP00,PN,Ono}.

\begin{figure}[t]
\centerline{
\includegraphics[width=.5\textwidth]{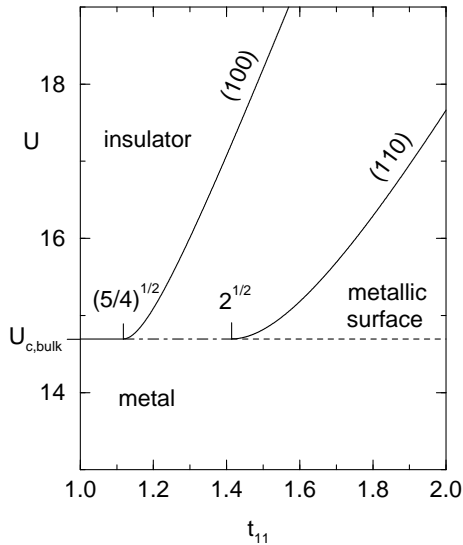}
}
\caption[]{
Phase diagram for the Mott transition at the (100) and (110) surface
of the sc lattice.
$t_{11}$ ($\ge t = 1$): modified hopping within the surface layer $\alpha=1$.
$U_{\rm c, bulk}$: bulk critical interaction.
}
\label{fig:results1}
\end{figure}

Within the L-DMFT the mean-field equations reduce to algebraic equations 
for $z_\alpha$ which involve the electronic model parameters and the system 
geometry.
For example, in the case of a surface geometry with $q_\|$ nearest 
neighbors within a layer, $q_\perp$ nearest neighbors within each of the 
adjacent layers and with uniform hopping $t$ and interaction $U$ except
for the hopping $t_{11} \ne t$ within the surface layer $\alpha = 1$,
the mean-field equations read:
\begin{eqnarray}
  z_1 = \frac{36}{U^2} 
               ( q_\| \, t_{11}^2 \, z_1 + q_\perp \, t^2 \, z_{2} ) 
\, , \quad
  z_\alpha = \frac{36 t^2}{U^2} 
             ( q_\| \, z_\alpha + q_\perp \, z_{\alpha+1} + q_\perp \, 
	     z_{\alpha-1} ) 
\nonumber \\
\label{eq:mf}
\end{eqnarray}
for $\alpha = 2,3,...,\infty$.
Equations of this type can be solved analytically or by simple numerical
means.

\begin{figure}[t]
\centerline{
\includegraphics[width=.55\textwidth]{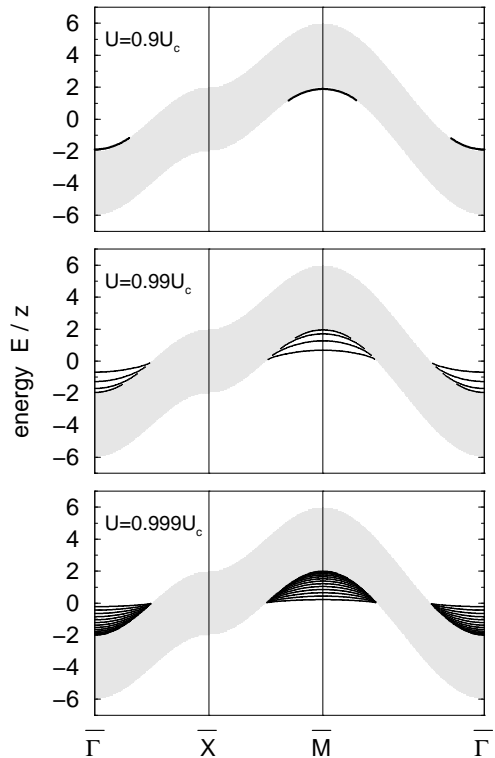}
}
\caption[]{
Surface states (lines) in the low-energy part of the one-electron excitation 
spectrum and bulk continuum (grey region) in the $D=2$ Brillouin zone for 
the unperturbed (100) surface ($t_{11} = t$) and different 
$U \to U_{\rm c, bulk}$.
Note that the energy scale is normalized with respect to the bulk 
quasi-particle weight $z$ with $z\to 0$ for $U\to U_{\rm c, bulk}$.
}
\label{fig:results2}
\end{figure}

Fig.\ \ref{fig:results1} shows the phase diagram for the Mott 
transition at two low-index surfaces of a semi-infinite $D=3$ simple cubic
lattice as obtained within the L-DMFT.
For the unperturbed surface ($t_{11} = t$) there is a single metal-insulator
transition at the bulk critical interaction (``ordinary transition'').
Within the L-DMFT $U_{\rm c, bulk} = 6t \: \sqrt{q_{\|} + 2 q_\perp} = 
6\sqrt{6}$.
An enhancement of the surface hopping $t_{11} > t$ which exceeds
a certain critical value, leads to two critical interactions:
At $U = U_{\rm c, bulk}$ there is the transition of the bulk of the system
irrespective of the state of the surface (``extraordinary transition'').
The surface undergoes its own phase transition to the insulating state at 
a second critical interaction
$U_{\rm c, surf} > U_{\rm c, bulk}$ (``surface transition'').
The critical perturbation $t_{11,\rm c}$ depends on the surface geometry.
Multi-critical behavior is found for $t_{11} = t_{11,\rm c}$ and 
$U = U_{\rm c, surf} = U_{\rm c, bulk}$ (``special transition'').

For $U_{\rm c, bulk} < U < U_{\rm c, surf}$ a metallic surface is coexisting
with an insulating bulk. 
As is demonstrated by Fig.\ \ref{fig:schema} this implies the existence of
a surface state.
Fig.\ \ref{fig:results2} shows that even for the unperturbed 
surface ($t_{11} = t$) in the metallic phase close to the ordinary transition
there is a surface state and in the limit $U \to U_{\rm c, bulk}$ even an 
infinite number of surface states.
These are split off from the bulk continuum of coherent low-energy excitations
the width of which vanishes proportional to $z \propto U_{\rm c, bulk} - U$
for $U \to U_{\rm c, bulk}$.
The critical profile of the quasi-particle weight in this limit, 
$z_\alpha \propto \alpha$, causes a strong surface perturbation of the 
low-energy electronic structure which drives the surface states.
These surface states should be considered as a new kind of 
{\em correlation-induced}
surface states as for the corresponding 
non-interacting model (with $t_{11}=t$) the occurrence of surface
states is impossible.

Concluding, one finds a phase diagram with the same topology as predicted
by the Landau theory of surface phase transitions.
This analogy can even be made quantitative: 
Consider Weiss mean-field theory for ferromagnetic order in the Ising model
on a semi-infinite lattice.
For $T \to T_{\rm C}$, the mean-field equation can be 
linearized, and one has 
$m_\alpha=(J/2T)(q_\|\, m_\alpha+q_\perp\, m_{\alpha+1}+q_\perp\, m_{\alpha-1})$
which by comparing with Eq.\ (\ref{eq:mf}) immediately yields the following
correspondences:
$m_\alpha \Leftrightarrow z_\alpha$, $J/2 \Leftrightarrow 36t^2$ and 
$T \Leftrightarrow U^2$ where $J$ is the coupling constant and $m_\alpha$ 
the layer-dependent magnetization.
It is obvious that this analogy has a number of implications for the Mott transition.

For example, for a thermodynamic phase transition (analogously for the $T=0$ 
Mott transition) there are two critical exponents that merely involve the 
critical temperature (interaction strength), the ``shift exponent'' $\lambda_s$
and the ``crossover exponent'' $\phi$ \cite{Bin83,Die86}. 
They describe the trend of $T_{\rm c}$ ($U_{\rm c}$) for films with thickness 
$d\mapsto \infty$ and the trend of $T_{\rm c, surf}$
($U_{\rm c, surf}$) for the semi-infinite system 
near the special transition, respectively.
Within the Laudau theory (linearized DMFT) one finds $\lambda_s=2$ and 
$\phi=1/2$.

\section{Conclusions}

Among different types of metal-insulator transitions at a surface of a 
single crystal, the correlation-driven Mott transition from a normal metal 
to a paramagnetic insulator is distinguished as it offers a comparatively 
simple route to a metallic surface phase of an insulating bulk.
Formally, this is expressed by the equivalence between the respective 
linearized mean-field approaches to the Mott quantum-phase transition 
and to the thermodynamic (magnetic) phase transition.
It should be stressed that the equivalence implies that all results of 
the Landau theory of surface phase transitions have a unique counterpart 
for the surface Mott transition.
This includes phase diagrams, critical profiles of the quasi-particle 
weight, critical exponents and other critical behavior.
In this way a comprehensive and consistent mean-field picture of the 
characteristics of the surface Mott transition is obtained.
\\

The author would like to thank R. Bulla and W. Nolting for discussions 
and collaborations.
The work has been supported by the Deutsche Forschungsgemeinschaft within
the Sonderforschungsbereich 290.


\begin{thebibliography}{10}

\bibitem{Mil71}
D.~L. Mills,
 Phys. Rev. B {\bf 3}, 3887 (1971).

\bibitem{Bin83}
K.~Binder,
 {\em {\rm In:} Phase Transitions and Critical Phenomena, Vol. 8},
 ed. by C. Domb and J. L. Lebowitz (Academic, London, 1983).

\bibitem{Die86}
H.~W. Diehl,
 {\em {\rm In:} Phase Transitions and Critical Phenomena, Vol. 10},
 ed. by C. Domb and J. L. Lebowitz (Academic, London, 1986).

\bibitem{DDN98}
P.~A.~Dowben, M.~Donath and W.~Nolting, editors,
 {\em Magnetism and electronic correlations in local-moment systems: 
 Rare-earth elements and compounds} (World Scientific, Singapore, 1998).

\bibitem{isihei}
T.~Kaneyoshi, I.~Tamura, and E.~F. Sarmento,
 Phys. Rev. B {\bf 28}, 6491 (1983);
K.~Binder and D.~P. Landau,
 Phys. Rev. Lett. {\bf 52}, 318 (1984);
C.~Tsallis and E.~F. Sarmento,
 J. Phys. C {\bf 18}, 2777 (1985);
F.~Aguilera-Granja and J.~L. Mor\'an-L\'opez,
 Phys. Rev. B {\bf 31}, 7146 (1985);
P.~J. Jensen, H.~Dreyss\'e, and K.~H. Bennemann,
 Surf. Sci. {\bf 269/270}, 627 (1992).

\bibitem{PP01}
R.~Pfandzelter and M.~Potthoff,
 Phys. Rev. B {\bf 64}, 140405 (2001).

\bibitem{MW66}
N.~D. Mermin and H.~Wagner,
 Phys. Rev. Lett. {\bf 17}, 1133 (1966).

\bibitem{Geb97}
F.~Gebhard,
 {\em The Mott Metal-Insulator Transition}
 (Springer, Berlin, 1997).

\bibitem{Lip}
R.~Lipowsky,
 Z. Phys. B {\bf 45}, 229 (1982); 
 J. Phys. A {\bf 15}, L195 (1982).

\bibitem{DS92}
S.~G. Davison and M.~Stre\'slicka,
 {\em Basic Theory of Surface States} (Clarendon, Oxford, 1992).

\bibitem{SN01}
R.~Schiller and W.~Nolting,
 Phys. Rev. Lett. {\bf 86}, 3847 (2001).

\bibitem{Hub}
J.~Hubbard,
 Proc. R. Soc. London A {\bf 276}, 238 (1963);
M.~C. Gutzwiller,
 Phys. Rev. Lett. {\bf 10}, 159 (1963); 
J.~Kanamori,
 Prog. Theor. Phys. (Kyoto) 
{\bf 30}, 275 (1963).

\bibitem{Mot90}
N.~F. Mott,
 {\em Metal-Insulator Transitions} (Taylor and Francis, London, 1990).

\bibitem{GKKR96}
A.~Georges, G.~Kotliar, W.~Krauth, and M.~J. Rozenberg,
 Rev. Mod. Phys. {\bf 68}, 13 (1996).

\bibitem{BCVJO}
R.~Bulla, T.~A. Costi, and D.~Vollhardt,
 Phys. Rev. B {\bf 64}, 045103 (2001);
J.~Joo and V.~Oudovenko,
 Phys. Rev. B {\bf 64}, 193102 (2001).

\bibitem{Eng63}
F.~Englert,
 Phys. Rev. {\bf 129}, 567 (1963).

\bibitem{MV89}
W.~Metzner and D.~Vollhardt,
 Phys. Rev. Lett. {\bf 62}, 324 (1989).

\bibitem{GK92}
A.~Georges and G.~Kotliar,
 Phys. Rev. B {\bf 45}, 6479 (1992).
 
\bibitem{Jar92}
M.~Jarrell,
 Phys. Rev. Lett. {\bf 69}, 168 (1992).

\bibitem{PN}
M.~Potthoff and W.~Nolting,
 Phys. Rev. B {\bf 59}, 2549 (1999);
 Euro. Phys. J. B {\bf 8}, 555 (1999);
 Phys. Rev. B {\bf 60}, 7834 (1999).

\bibitem{CK94}
M.~Caffarel and W.~Krauth,
 Phys. Rev. Lett. {\bf 72}, 1545 (1994).

\bibitem{BP00}
R.~Bulla and M.~Potthoff,
 Euro. Phys. J. B {\bf 13}, 257 (2000).

\bibitem{Ono}
Y.~\=Ono, R.~Bulla, and A.~C. Hewson,
 Euro. Phys. J. B {\bf 19}, 375 (2001);
Y.~\=Ono, R.~Bulla, A.~C. Hewson, and M.~Potthoff,
 Euro. Phys. J. B {\bf 22}, 283 (2001).

\end{thebibliography}
\end{document}